\documentclass[conference]{IEEEtran}

\usepackage[english]{babel}
\usepackage{courier}
\usepackage{amstext}
\usepackage{times}
\usepackage{helvet}
\usepackage{courier}
\usepackage[cmex10]{amsmath}
\usepackage{amssymb}
\usepackage{latexsym}
\usepackage{eurosym}
\usepackage{wasysym}

\makeatletter
\newif\if@restonecol
\makeatother

\usepackage{etex}
\usepackage{tikz}
\usepackage{pslatex}
\usepackage[arrow, matrix, curve]{xy}
\usepackage{epsfig}
\usepackage{pstricks,egameps}
\usepackage{hyperref, url}

\begin{document}

\title{The Economy of Internet-Based Hospitality Exchange}

\author{\IEEEauthorblockN{Rustam Tagiew}
\IEEEauthorblockA{Alumni of\\TU Bergakademie Freiberg\\
yepkio@mail.ru}}

\maketitle

\begin{abstract}
In this paper, we analyze and compare general development and individual behavior on two non-profit internet-based hospitality exchange services -- bewelcome.org and warmshowers.org. We measure the effort needed to achieve a real-life interaction, whereby the advantages of mutual altruism arise. The effort needed is the communication quantified in units of time. Since the amount of effort is not obvious to individual users, the development of the effort investing strategy is investigated. The impact of individual behavior on general development is discussed.    
\end{abstract}


\IEEEpeerreviewmaketitle

\section{Introduction}

\indent The rite of gratuitous hospitality provided by local residents to unheralded strangers is common among almost all known human cultures, especially traditional ones. We can cite melmastia of Pashtun \cite[p. 14]{melmastia}, terranga of Wolof \cite[p. 17]{teranga} and hospitality of Eskimo \cite{eskimo} as examples. A stranger hereby receives shelter, food, protection and other kinds of help. In return, the guest is expected to contribute at least symbolically to his hosts well-being. According to numerous travel blogs and articles \cite[e.g.]{joinville}, this is still common today.\\
\indent In the modern world the share of tourists among the strangers has grown. A small share of tourists, which can be as {\it travelers}, seek for gratuitous hospitality from local residents. The advantage for travelers is not only reduction in monetary expenses \cite{anja}, but also engagement in relationships shaped by altruism termed as {\it social market} \cite{twomarkets}. Participants of a social market can obtain for free services and goods affordable or unavailable for them on monetary market. For instance, a budget traveler can receive an accommodation equivalent to a 3-star hotel room while their host enjoys a pleasant soir\'ee unsellable anywhere else.\\
\indent Because travelers can turn into local residents and vice versa, the idea of hospitality exchange abbreviated as \textit{hospex} became relevant. Hospex participants host and can have hosts. They are organized in a community, where gratuitous hospitality is not responded directly but by hospitality of another participant or rather member. In the long run, accommodation does not remain the only type of real-life interactions, which provide the advantages of a social market. Further, we will term everybody as {\it member}, who had a real-life interaction with another member.\\
\indent For a functioning hospex community, a {\it hospex service} is needed. It basically administrates so-called {\it profiles} -- participant description, contact and location. A hospex service is easier to maintain, once a central online database of profiles is set up. In addition to profiles, recording accommodation reports aka references or {\it comments} became popular for online hospex services. Such a database is accessible over a web- and/or app-based front-end. At first, such service called 'Hospex' was created in 1992 \cite{hospex}. The installations of other services followed. Hospitalityclub.org abbreviated as HC became the biggest of them with over 100k profiles in 2006 \cite{hcpaper}. In the years 2006-2009, a user migration from HC to Couchsurfing.org abbreviated as CS took place \cite{bewelcomepaper}. CS has now the biggest number of profiles -- over 3M \cite{cssite}. We have to underline here the difference between the terms \textit{user}, which is basically a profile owner, and \textit{
member} -- not every website user had a real-life interaction with other members in order to be called a member. Members are a subset of users.\\ 
\indent Unfortunately, we can not access the data from CS -- the biggest hospex service. CS became a for-profit corporation in 2011 and shut down the access of public science to its data. Scientists possessing pre-incorporation CS data are prohibited to share it with third parties. HC never allowed access of public science to its data. Therefore, we can only rely on the data kindly mirrored by Bewelcome.org (BW) on 04.03.2014 and Warmshowers.org (WS) on 03.01.2015. BW is meant for general hospex users and WS specializes on bicycle travelers. Both are run by legally non-profit organizations.\\  
\indent Hospex services obviously satisfy inter alia the need, which hotel business is based on. Having an impact on economy hereby, hospex services are seen as its part -- the so-called {\it share economy} \cite[e.g.]{sharecs}. The term 'share economy' suggest that the common economy is non-share -- it originates from Weitzmans work proposing share economy as a cure to macroeconomic problems \cite{shareeco}. 'Share economy' is now a widely used term far away from a clear scientific definition.\\
\indent Non-profit hospex services are a special case -- money does not play any role. We have to augment the long list of diverse definitions of economy and economics \cite{defecon} in order to include hospex. For this research, economy is defined as the set of activities in a human community, by what effort results in benefit. Economics is the study of rationality in economies. Economics replies the question, how effort can be sustainably reduced and benefit sustainably increased. Relying on this definition, we have to measure the effort and the benefit for individuals participating in hospex through internet. The benefit are obviously real-life interactions, whereby the advantages of mutual altruism arise. The effort might be the communication, which is to be measured in units of time. The question to be answered is, whether individuals adjust the amount of their effort and how does that impact general development.\\
\section{Related Work}
\indent We can identify three categories of existing scientific publications claiming hospex as their subject matter -- non-data scientific articles, analysis of survey data, and analysis of CS data. Survey data gives us insights into mindsets, but not into real behavior processes on hospex. Currently we know about four research teams, who used pre-incorporation CS data \cite{victor,danderkar,overgoor,lauterbach}. All papers written by these research teams solely concentrate on the aspect of trust among the CS users. They don't particularly investigate economical aspects as aimed in this paper. Further, the correctness of their work can not be double-checked, because they are not allowed to share the data anymore. An observational study on BW data has already been published \cite{bewelcomepaper}.\\
\indent This work combines knowledge and methods behavioral of economics and data science, where market leaders already push forward into. Facebook hired economists for studying the behavior of users and advertisers, economic networks, incentives, externalities, and decision making under risk and uncertainty \cite{facebookbe}. Microsoft founded MSR-NYC \cite{microsoftbe}, where researchers develop technologies in the intersection of social science, both computational and behavioral, computational economics and prediction markets, machine learning, as well as information retrieval. Google's chief economist openly writes in his paper \cite{varian}: ``I believe that [manipulating and analyzing big data] have a lot to offer and should be more widely known and used by economists. In fact, my standard advice to graduate students these days is `go to the computer science department and take a class in machine learning'.''\\
\indent In the academic world, the international workshop series ``Experimental Economics and Machine Learning'' (EEML) started in 2012 \cite{eeml1be,eeml2be,eeml3be}. EEML seeks to fill the gap between two scientific communities of Experimental Economics and AI \& Data Mining. The conference ``Social media and behavioral economics'' took place in 2013 \cite{harvardbe}, where data scientists and economists from universities and industry participated. Yale University has a chair researching in machine learning, behavioral economics, and finance \cite{brown}. The term 'Behavior Mining' is suggested for the analysis of human behavior from web data, whereby the semantics of complex data  is injected into the mining process \cite{Chen06}.
\section{General Behavior}
\begin{figure}
\includegraphics[scale=0.5]{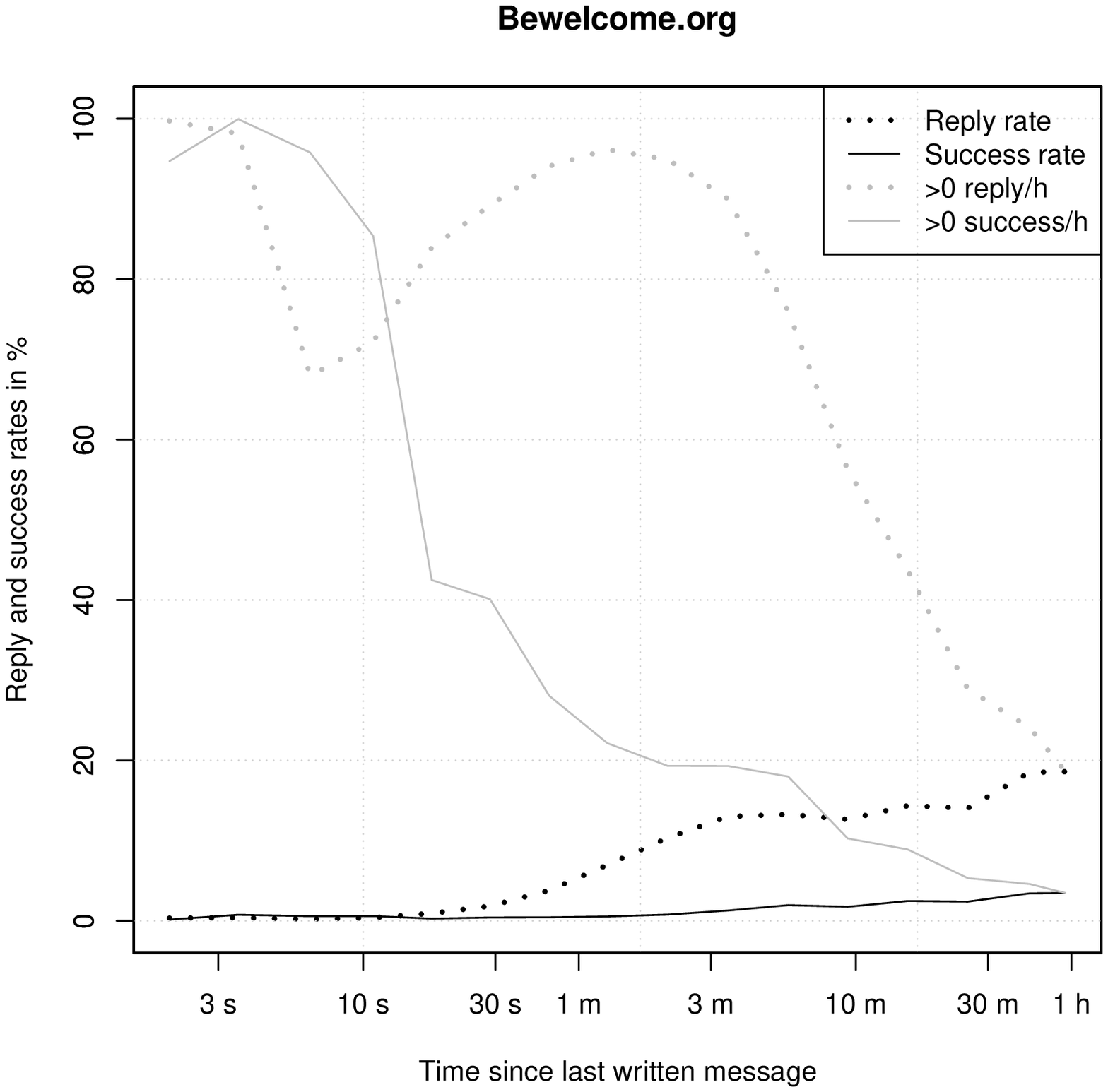}
\includegraphics[scale=0.5]{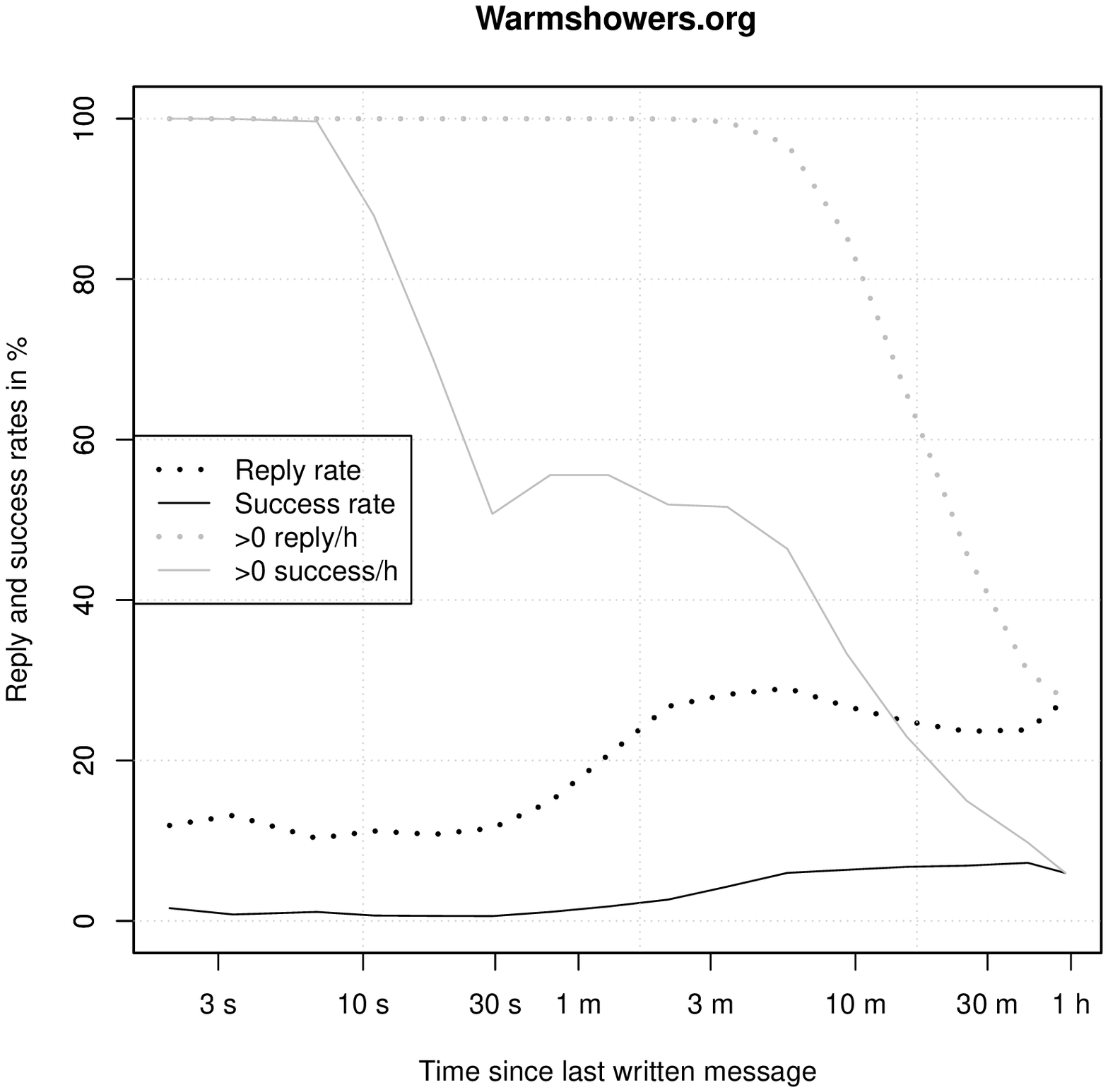}
\caption{Black lines plot histograms of success and reply rates over $20$ logarithmically distributed bins. Grey lines plot theoretical probabilities for at least one reply or success respectively if sending an hour writing initiations of certain time consumption.}
\label{successint}
\end{figure}

\begin{figure}
\includegraphics[scale=0.5]{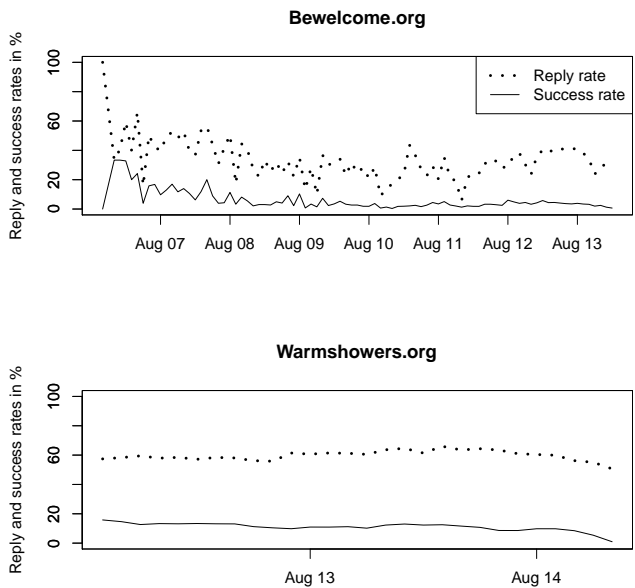}
\caption{Dependence of reply and success rates on month. BW had only a tiny fraction of todays user number in the year 2007--2008. The little drop in last months is due to time delay between writing an initiation and receiving/writing a comment.}
\label{successmm}
\end{figure}

\begin{table}
\begin{center}
\caption{Reply rate, speed and success rate on BW and WS.}
\label{replyonhospex}
\begin{tabular}{|l|c|c|}
\hline
\hline
                                                        &  BW         & WS       \\
\hline
Replied initiations                                    &  $31.8$\%   & $61.6$\% \\
\hline
Share of first replies arriving within 24 hours         &  $60$\%     & $65.2$\% \\
\hline
Share of first replies arriving within 7 days           &  $86.8$\%   & $90.5$\% \\
\hline
Initiations resulting in a real-life interaction        &             &          \\
$\equiv$ Success rate                                   &  $3.7$\%    & $10.5$\% \\
\hline
\end{tabular} 
\end{center} 
\end{table}
\indent Since the data is complex and incomplete, many reasonable assumptions based on context knowledge have to be met in order to conduct its analysis. Let us term a message exchange between two users as a {\it conversation}. The first message in a conversation is an {\it initiation} and comes from an {\it initiator} and the other user is a {\it follower}. Before a conversation, users are generally assumed to be strangers to each other. An initiation is assumed to be a request sent from a traveler to a potential host in most of the cases. It can also be an invitation, a briefing about a gathering or even just a question.\\
\indent Not every initiated conversation is followed within a hospex service. $56.6$\% of total $895042$ messages from WS data and $66.3$\% of total $227690$ messages from BW data are initiations. There are cases, where initiator prefers staying unanswered -- mass-sending of general information like welcoming new users e.g.. The follower is expected to respect initiators preferences. Since the messages are neither pre-categorized in BW data nor in WS data, the exclusion of these cases from evaluation requires the application of approximative automatic categorization of message text, which is not available due to privacy policies at least for BW. This procedure is not performed.\\
\begin{figure}
\includegraphics[scale=0.5]{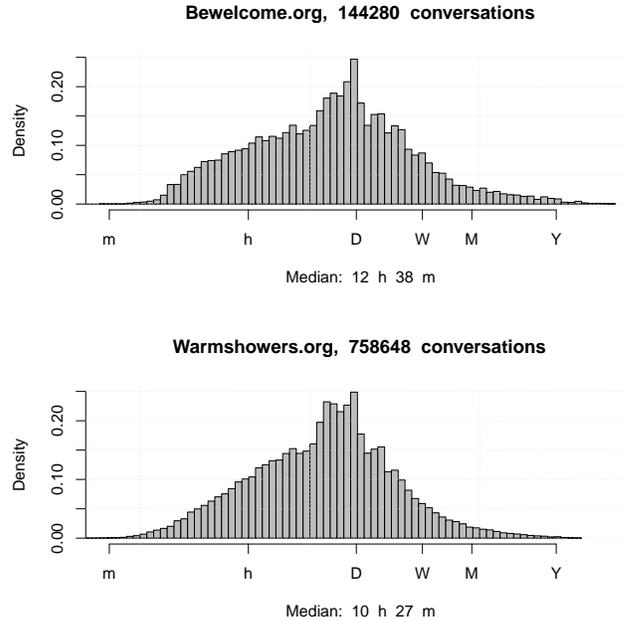}
\caption{Time intervals between initiations and replies for replied initiations. Time axis is logarithmic: m - minute, h - hour, D - day, W - week, M - month and Y -year.}
\label{replyplot}
\end{figure}
\indent Tab.\ref{replyonhospex} displays a strong difference in rates of replied initiations termed as {\it reply rates} between BW and WS. We assume that included mass-sending of general information is not the major reason for it. If an initiation is replied on BW or WS, the majority of replies arrives within $24$ hours. BW and WS show visually similar distributions for reply latencies (Fig.\ref{replyplot}). The time axis of both plots are chosen to be logarithmic for better visualization. A conversation can be continued using other media like mobile phones, e-mails and so on. Some users openly publish contact details on their profiles -- conversations can be realized even without using particular hospex messaging feature. There are indeed cases of comments being written between users, where no recorded conversations exist.\\ 
\begin{table}
\begin{center}
\caption{Message writing speed on hospex services.}
\label{mwritinghospex}
\begin{tabular}{|l|c|c|}
\hline
\hline
First cluster of interval distribution                  &  BW         & WS       \\
\hline 
$\mu$, log milliseconds                                 &  $11.535500$ & $12.064600$ \\
$\sigma$, log milliseconds                              &  $1.800600$  & $1.432200$ \\
cluster size                                            &  $107731$    & $349829$ \\  
\hline
$e^{\mu+2*\sigma}\approx$ $97.5$\% upper bound          &  $1$ h $2$ min $27$ s & $50$ min $44$ s \\
\hline
\end{tabular} 
\end{center} 
\end{table}
\begin{figure}
\includegraphics[scale=0.5]{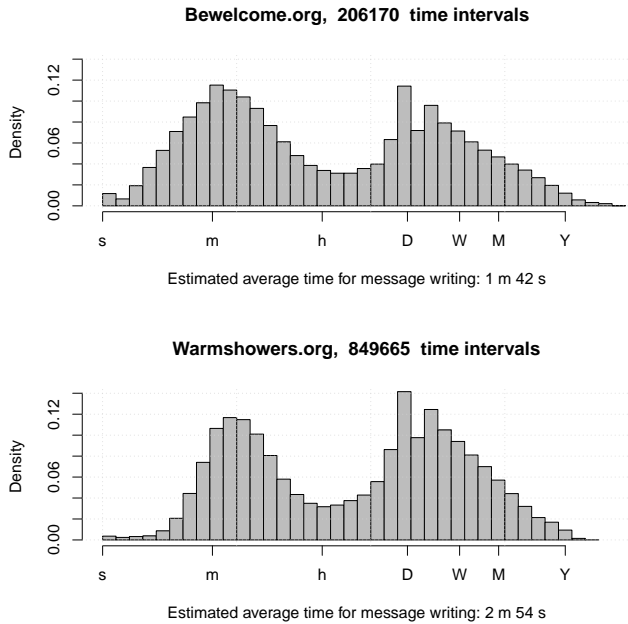}
\caption{Time intervals between subsequent sent messages. Time axis is logarithmic: s - second, m - minute, h - hour, D - day, W - week, M - month and Y -year.}
\label{messageintervals}
\end{figure}
\indent If an comment is written by a user for a user, we assume that both had most probably a real-life interaction. Otherwise, we assume the number of real-life interactions leading to no comments being written from both sides to be minimal. Although not every user pair involved in a conversation intends to meet in real-life, real-life interactions are regarded as the general goal of communication through hospex services \cite{anja}. Therefore we term the proportion of initiations leading into a real-life interaction as {\it success rate}. The overall success rates for both services are displayed on Tab.\ref{replyonhospex}.\\   
\indent Considering the relatively low success rate on both hospex services, the amount of communication respectively effort needed to achieve at least one real-life interaction is significant. In order to estimate that effort, intervals between subsequent messages are calculated. This is only possible for users, who sent more than one message in total. Fig.\ref{messageintervals} shows the distribution of the intervals for both services. The distributions are obviously similarly shaped and have two major bell-formed hills -- mixed log-normal distribution. Log-normal distribution was already observed in other human communication data \cite{lognormal}. The distribution for reply intervals on Fig.\ref{replyplot} is analogously hypothesized to be complex mixed log-normal. The first hill for BW and WS on Fig.\ref{messageintervals} obviously refers to messages being written uninterrupted by any other activity and the second one for the rest of messages. One dimensional EM-clustering fixed on cluster number of $2$ 
yields means and standard deviations of both hills. Tab.\ref{mwritinghospex} displays the results for the first cluster. \\
\begin{table}
\begin{center}
\caption{Initiation speed as bias for reply and success rates.}
\label{initspresuc}
\begin{tabular}{|l|c|c|}
\hline
\hline
First cluster of interval distribution                  &  BW          & WS       \\
\hline 
$\mu$, log milliseconds                                 &  $11.1784$   & $11.8773$ \\
$\sigma$, log milliseconds                              &  $1.6695$    & $1.32$ \\
cluster size                                            &  $62529$     & $109887$ \\  
\hline
$e^{\mu+2*\sigma}\approx$ $97.5$\% upper bound          &  33 m 38 s   & 33 m 37 s \\
\hline
Inside 1 h upper bound                                  &              &           \\
Success rate                                            &  $1.1$\%     & $1.8$\% \\
Reply rate                                              &  $7.9$\%     & $11.7$\% \\
\hline
Outside 1 h upper bound                                 &              &           \\
Success rate                                            &  $6.9$\%     & $14.2$\% \\
Reply rate                                              &  $61$\%      & $80.6$\% \\
\hline
Inside 1 h upper bound, log-normal means                &            &        \\
Replied initiations                                     &  $3$ m $56$ s & $3$ m \\
Unreplied initiations                                   &  $1$ m $2$ s  & $2$ m $7$ s \\
Real-life interaction triggering                        &  $3$ m $10$ s & $5$ m $2$ s \\  
No real-life consequences                               &  $1$ m $8$ s      & $2$ m $14$ s \\  
\hline
\end{tabular} 
\end{center} 
\end{table}
\begin{figure}
\includegraphics[scale=0.5]{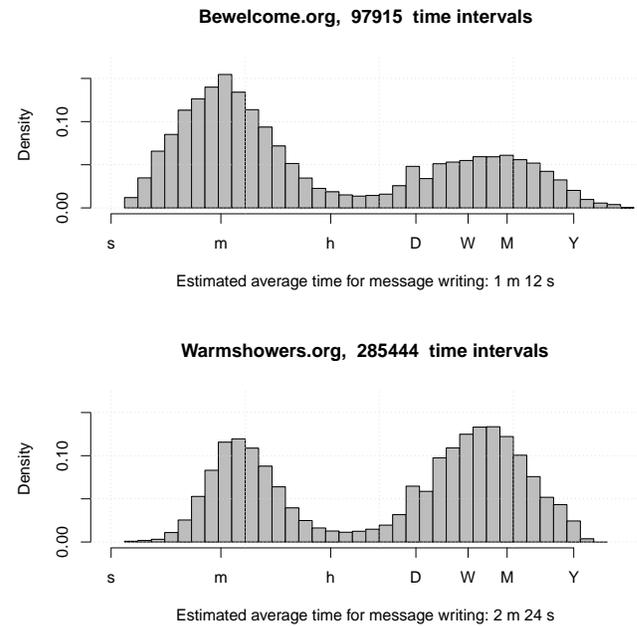}
\caption{Time intervals between subsequent sent conversation intiations. There are less of those intervals than conversation, because first messages have no previous message. Time axis is logarithmic: s - second, m - minute, h - hour, D - day, W - week, M - month and Y -year.}
\label{initintervals}
\end{figure}
\indent Fig.\ref{initintervals} depicts the distributions of intervals since last written messages for initiations. The distributions for BW and WS strongly differs here. $63.9$\% of intervals ahead of initiations on BW belong into first cluster and only $38.5$\% on WS. BW users send relatively more subsequent initiations than WS users. Such set of initiations to term as {\it bundle} increases the probability to achieve at least one success according to Eq.\ref{eq1}.
\begin{equation}\label{eq1}
P_{success>0}(x) = 1-{(1-p_{success})}^x\text{, x -- bundle size}
\end{equation}
Sending bundles is only possible, if more than one potential host has indicated to inhabit a certain location. Such location are the major cities. Unlike dependent and motorized travelers staying at major cities, cyclists on WS have to pass nights in rural areas with only one potential host per location if at all.\\
\indent A comparison of Tab.\ref{mwritinghospex} and Tab.\ref{initspresuc} shows that log-normal means for writing initiations are significantly lower than for other messages on BW as well as on WS. Log-normal standard deviations are lower as well. As result, $97.5$\% upper bound for initiations is roughly $\tfrac{1}{2}$ hour for both hospex services, while it is the double for general messages.\\
\indent Reply and success rates grow with time assumed to be spent for writing initiations for both investigated hospex services. Tab.\ref{initspresuc} already displays that the initiations written within an hour since last message have dramatically lowered reply and success rates. Average replied and successful initiations within an hour since the last message take significantly more time than average unreplied and unsuccessful ones. On Fig.\ref{successint}, histograms of reply and success rates over $20$ logarithmically distributed bins are depicted and confirm this observation. The number of bins is arbitrarily chosen. Visually, reply and success rates of initiation are positively correlated with time consumption for their production. Success rates are even stronger correlated than reply rates -- BW at $0.91$ and WS at $0.705$ having $20$ bins. Either users are taking more time to write initiations more likely to be replied and have real-life interactions or more time spent on message writing improves the reply 
and success rates independently of user. We assume the first case to be negligible, since a user's only way to impact the reply and success rates except by message text is h(er)is profile, which is written with on-line as well. Reply and success rates do not change over time as shown on Fig.\ref{successmm}.\\
\indent The explanation for the growing of success rate with effort per initiation is most likely sending a wrong signal. Since the interaction on hospex services are shaped by altruism and effort does not depend on benefit in altruistic relationships \cite{twomarkets}, an attempt to reduce the effort from the start on would confuse the follower about the real intentions of the initiator.\\
\begin{figure}
\includegraphics[scale=0.5]{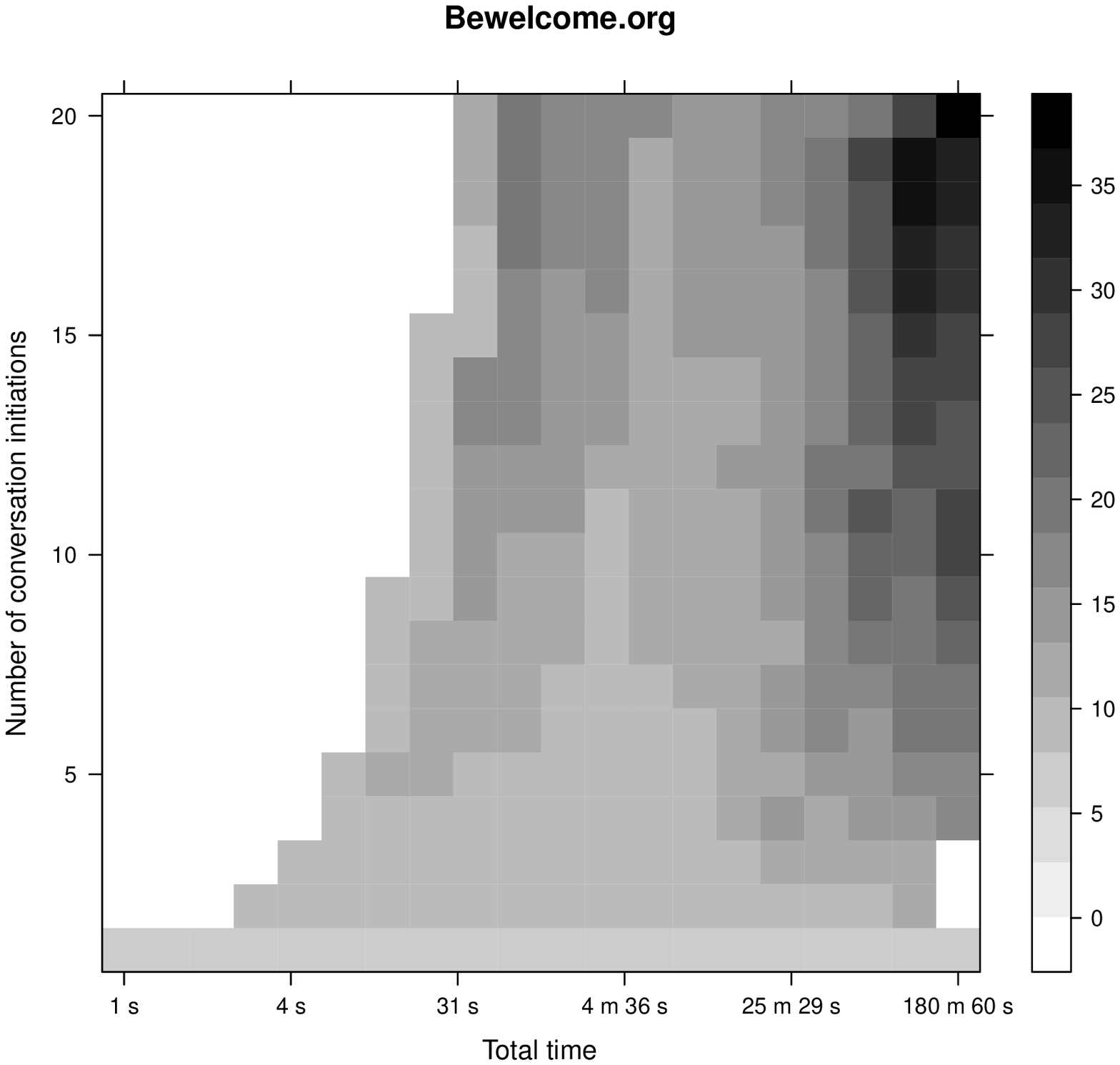}
\includegraphics[scale=0.5]{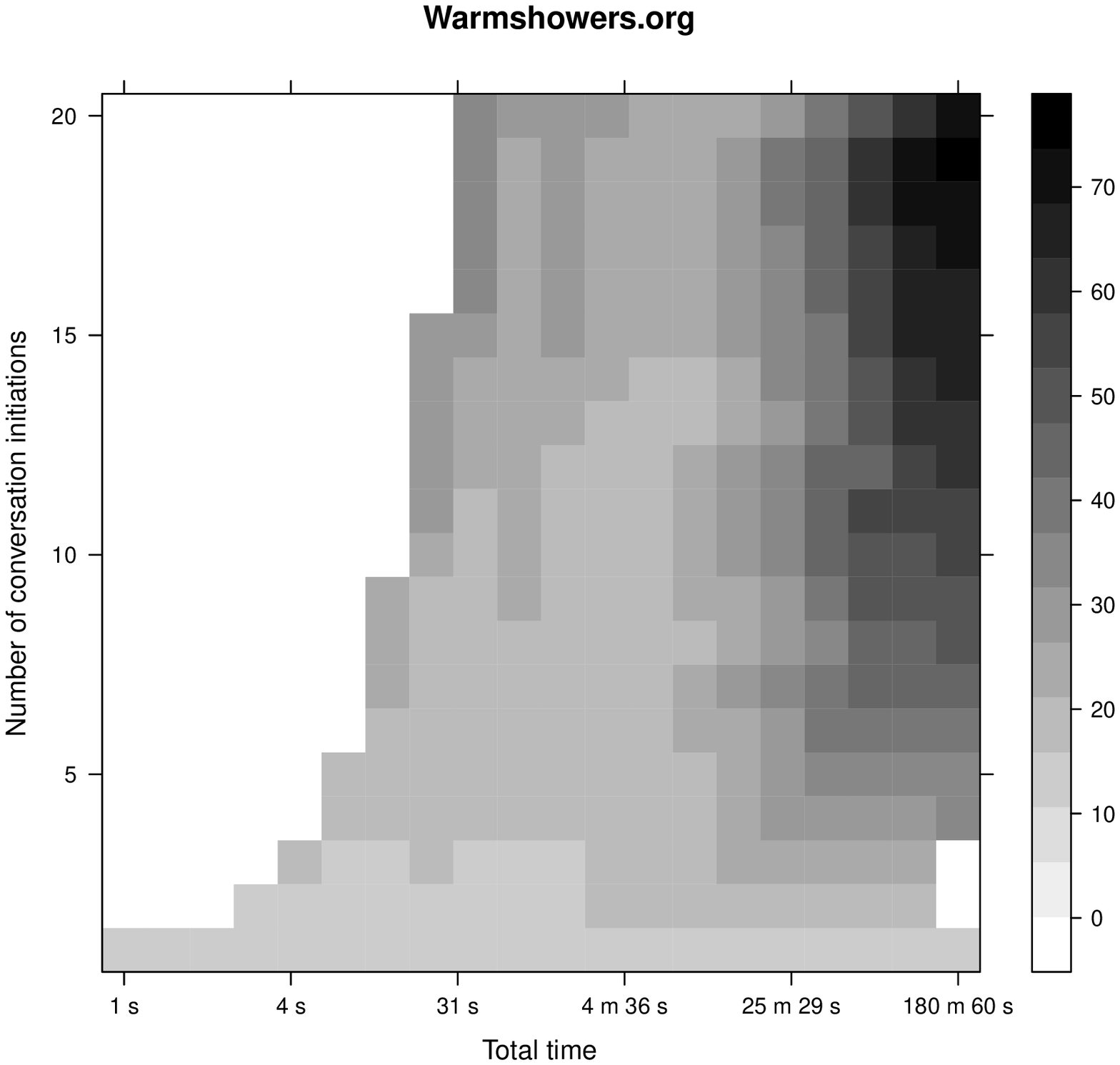}
\caption{Theoretical probability for at least one real-life interaction depending on total time and number of sent initiations in \%.}
\label{successlat}
\end{figure}
\indent Success rate $p_{success}$ in Eq.\ref{eq1} is a function of time spent for single initiation. Therefore the probability for at least one reply $P_{success>0}$ after sending a bundle is a function of time and bundle size. Grey lines on Fig.\ref{successint} depict the plots of theoretical $P_{success}$ having total time of an hour. Although less time spent on a single initiation reduces its success rate, it increases the number of initiations being sent per time unit. In result, quantity outperforms quality if of course a big number of followers is available to send initiations to. Nevertheless, sending many fast written initiations -- the reportedly obnoxious 'copy and paste requests' \cite{candpcs} -- is not as reasonable for small BW and WS as for big CS with huge amounts of potential hosts per location. Fig.\ref{successlat} shows the plot of $P_{success>0}$ as a function of total time and bundle size for both hospex services. According to this plot, if you have limited time, you should send initiations to 
as many as possible, and if you have limited number of followers, more time per initiation increases the probability to have at least one real-time interaction.\\ 
\section{Individual Behavior}
\begin{table}
\begin{center}
\caption{Experience and time spent on one initiation.}
\label{hypothesis}
\begin{tabular}{|l|c|c|}
\hline
\hline
                                                        &  BW          & WS       \\
\hline 
Number of instances                                     &  $12898$     & $13316$  \\
\hline
Correlation of initiation effort with ...                        &              &          \\  
(P-values) ...                        &              &          \\  
... average initiation effort $<$1h & $ 0.435 $ ($ 0 $)   & $ 0.49 $ ($ 0 $) \\
... number of initiations & $ -0.044 $ ($ 0 $)    & $ -0.155 $ ($ 0 $) \\
... reply rate & $ 0.258 $ ($ 0 $)      & $ 0.123 $ ($ 0 $) \\
... success rate & $ 0.268 $ ($ 0 $) & $ 0.196 $ ($ 0 $) \\
... avg. successful initiation effort $<$1h $\equiv$ ASI & $ 0.206 $ ($ 0 $)     & $ 0.252 $ ($ 0 $)\\
... avg. unsuccessful init. eff. $<$1h $\equiv$ AUI & $ 0.419 $ ($ 0 $)  & $ 0.441 $ ($ 0 $)\\
... ASI - AUI & $ 0.023 $ ($ 0.01 $) & $ 0 $ ($ 0.992 $)   \\
... ASI/AUI & $ -0.014 $ ($ 0.107 $) & $ 0.011 $ ($ 0.224 $) \\
\hline
\end{tabular} 
\end{center} 
\end{table}
\indent The function of $P_{success>0}$ in dependence of total time and bundle size is likely to be not obvious for the worm's-eye view of a user from the start on. Our hypothesis is that users try to increase their success rate and learn from their experience. In the ideal case, users should notice that more effort spent for an initiation results in more success. And since the BW and WS are small, users would increase the effort per initiation rather than the size of a bundle. In case, where user experience contradicts due to the randomness of the global trend, (s)he will reduce the effort per initiation. This means that the time spent on an initiation should correlate with the experienced relationship of effort to benefit.\\
\indent Unfortunately, we have to reject our hypothesis after the evaluation of single human decisions regarding the extent of depending on effort experience. Tab.\ref{hypothesis} displays that the effort correlates the most with the average of own previous effort from the set of selected features. Since it correlates with the average of own previous effort, it correlates with the averages of own previous effort for un- and successful initiations at a lower level as well. It correlates with reply and success rates, since users spending more time per initiation are known to achieve more replies and real-life interactions per initiation. There is even a slight but significant negative correlation between effort and number of already written initiation -- a slight reduction of time needed per initiation most probably due to the increase in professionalism. But, there is no or almost no correlation between effort and the experienced relationship of effort to benefit. Users either do not learn or don't change 
their behavior as result of their acquired knowledge.\\
\newpage
\section{Conclusion}
\indent A hospex community like any other community is economically reasonable, if the effort invested does not exceed the benefit received. Nevertheless, since a hospex community is shaped by altruism, low-effort communication sends the wrong signal and reduces the success rate per initiation. That makes low-effort communication also known as 'copy and paste requests', whereby quality is replaced by quantity, only reasonable for hospex services with a big user base like CS. The BW success rate is approximately half of WS, if many assumptions about incomplete data made earlier are true. BW is not in the same league as WS, since WS specializes on bicycle travelers. Individual users do not adjust the extent of their effort for writing initiations according to their experience. Since more than one hospex service is available, individuals can migrate to ones with the higher benefit per effort ratio. One can hypothesize that the impossibility of low-effort communication on BW makes CS appear more attractive for the majority of travelers. Anyway, todays budget travelers can not afford to abstain from hospex services -- any monetary consideration of effort and benefit would result a very high hourly pay rate.\\   
\section*{About and Acknowledgment}
\indent This work was partially published on Bewelcome.org and Warmshowers.org under a Creative Commons License \cite{bwdr}. I want to express my gratitude to the co-members of Bevolunteer.org and Randy Fay, the founder of WS, for mirroring the data.\\
\bibliographystyle{IEEEtran}
\bibliography{webmine01}
\end{document}